\documentclass[submission]{ieee}

\usepackage{amsmath}
\usepackage{amsfonts}
\usepackage{amssymb}

\usepackage{graphicx}

\begin{document}

\title[Measurement of the sensitivity function in ...]{%
       Measurement of the sensitivity function in time-domain atomic interferometer}
\author[CHEINET et al.]{P. Cheinet, B. Canuel, F. Pereira Dos Santos, A. Gauguet, F. Leduc, A. Landragin
\authorinfo{P. Cheinet, B. Canuel, F. Pereira Dos Santos, A. Gauguet, F. Leduc and A.
Landragin are with Laboratoire BNM-SYRTE,75014 Paris, France
(e-mail: patrick.cheinet@obspm.fr)} }

\date{MARCH 24, 2005}

\journal{IEEE Trans.\ on Instrum.\ Meas.}
\loginfo{Manuscript received }
\firstpage{1}

\maketitle

\begin{abstract}
We present here an analysis of the sensitivity of a time-domain
atomic interferometer to the phase noise of the lasers used to
manipulate the atomic wave-packets. The sensitivity function is
calculated in the case of a three pulse Mach-Zehnder
interferometer, which is the configuration of the two inertial
sensors we are building at BNM-SYRTE. We successfully compare this
calculation to experimental measurements. The sensitivity of the
interferometer is limited by the phase noise of the lasers, as
well as by residual vibrations. We evaluate the performance that
could be obtained with state of the art quartz oscillators, as
well as the impact of the residual phase noise of the phase-lock
loop. Requirements on the level of vibrations is derived from the
same formalism.

\end{abstract}



\section{Introduction}
\begin{figure}[h]
\centering
\includegraphics[width=5in]{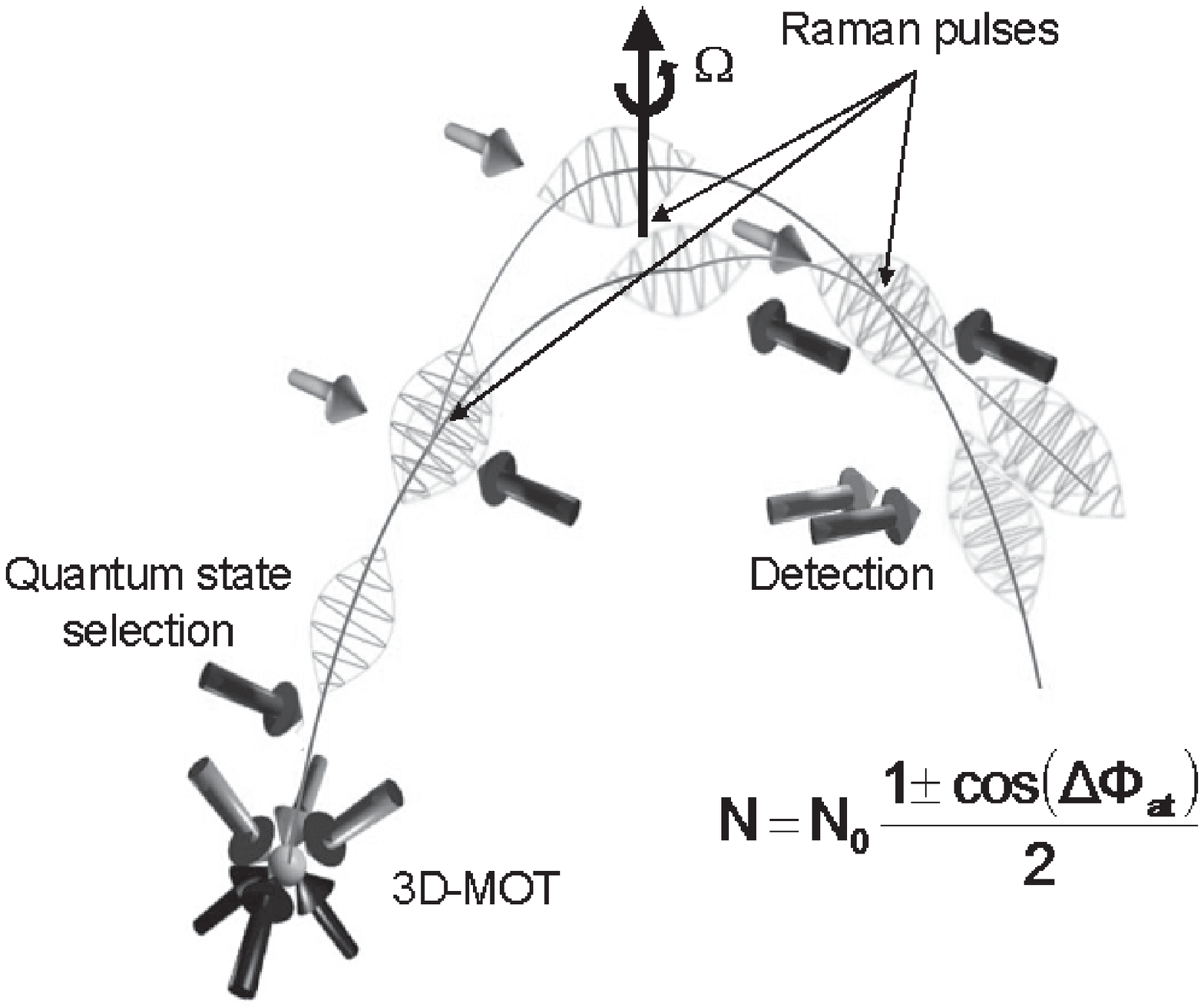}
\caption{Scheme of principle of our inertial sensors, illustrated for the gyroscope experiment.
Cold atoms from the 3D-MOT are launched upwards and a pure quantum state is selected. At the top of
their trajectory, we apply three Raman laser pulses realizing the interferometer. Finally a
fluorescence detection allows to measure the transition probability. Such an interferometer is
sensitive to the rotation ($\Omega$) perpendicular to the area enclosed between the two arms and to
the acceleration along the laser's axis.} \label{gyro}
\end{figure}

\PARstart{A}{tom} optics is a mean to realize precision
measurements in various fields. Atomic microwave clocks are the
most precise realization of a SI unit, the second
\cite{Clairon95}, and high sensitivity inertial sensors
\cite{Riehle91,Gustavson00,Peters01}, based on atomic
interferometry \cite{Berman97}, already reveal accuracies
comparable with state of the art sensors
\cite{Niebauer95,Stedman97}. Two cold atom inertial sensors are
currently under construction at BNM-SYRTE , a gyroscope
\cite{Leduc03} which already reaches a sensitivity of
$2.5\times10^{-6}\,\rm{rad.s^{-1}.Hz^{-1/2}}$, and an absolute
gravimeter \cite{Cheinet01} which will be used in the BNM Watt
Balance project \cite{Geneves05}. Although based on different
atoms and geometries, the atomic gyroscope and gravimeter rely on
the same principle, which is presented in figure \ref{gyro}. Atoms
are collected in a three dimensional magneto-optical trap (3D-MOT)
in which the atoms are cooled down to a few ${\mu} K$. In the
gyroscope, $^{133}$Cs atoms are launched upwards with an angle of
8\r{} with respect to verticality using the technic of moving
molasses, whereas in the gravimeter, $^{87}$Rb atoms are simply
let to fall. Then the initial quantum state is prepared by a
combination of microwave and optical pulses. The manipulation of
the atoms is realized by stimulated Raman transition pulses
\cite{Kasevich91}, using two counter-propagating lasers, which
drive coherent transitions between the two hyperfine levels of the
alkali atom. Three laser pulses, of durations
$\tau_R-2\tau_R-\tau_R$, separated in time by $T$, respectively
split, redirect and recombine the atomic wave-packets, creating an
atomic interferometer \cite{Borde91}. Finally, a fluorescence
detection gives a measurement of the transition probability from
one hyperfine level to the other, which is given by
$P=\frac{1}{2}(1-\cos(\Phi))$, $\Phi$ being the interferometric
phase. The phase difference between the two Raman lasers (which we
will call the Raman phase throughout this article, and denote
$\phi$) is printed at each pulse on the phase of the atomic wave
function \cite{Borde03}. As $\phi$ depends on the position of the
atoms, the interferometer is sensitive to inertial forces, and can
thus measure rotation rates and accelerations. A drawback of this
technic is that the measurement of the interferometric phase is
affected by the phase noise of the Raman lasers, as well as
parasitic vibrations. The aim of this article is to investigate
both theoretically and experimentally how these noise sources
limit the sensitivity of such an atomic interferometer.

\section{sensitivity function\label{sec:sensitivity function}}

The sensitivity function is a natural tool to characterize the
influence of the fluctuations in the Raman phase $\phi$ on the
transition probability \cite{Dick87}, and thus on the
interferometric phase. Let's assume a phase jump $\delta \phi$
occurs on the Raman phase $\phi$ at time t during the
interferometer sequence, inducing a change of $\delta P(\delta
\phi,t)$ in the transition probability. The sensitivity function
is then defined by :

\begin{equation}
\label{eq} \ g(t)=2 \lim_{\delta \phi\rightarrow 0} \frac{\delta
P(\delta \phi,t)}{\delta \phi }.\
\end{equation}

The sensitivity function can easily be calculated for
infinitesimally short Raman pulses. In this case, the
interferometric phase $\Phi$ can be deduced from the Raman phases
$\phi_1$,$\phi_2$,$\phi_3$ during the three laser interactions,
taken at the position of the center of the atomic wavepacket:
$\Phi=\phi_1-2\phi_2+\phi_3$ \cite{Kasevich92}. Usually, the
interferometer is operated at $\Phi=\pi/2$, for which the
transition probability is 1/2, to get the highest sensitivity to
interferometric phase fluctuations. If the phase step $\delta
\phi$ occurs for instance between the first and the second pulses,
the interferometric phase changes by $\delta\Phi=-\delta\phi$, and
the transition probability by $\delta P=-cos(\pi/2+\delta \Phi)/2
\sim -\delta \phi/2$ in the limit of an infinitesimal phase step.
Thus, in between the first two pulses, the sensitivity function is
-1. The same way, one finds for the sensitivity function between
the last two pulses : +1.

In the general case of finite duration Raman laser pulses, the
sensitivity function depends on the evolution of the atomic state
during the pulses. In order to calculate $g(t)$, we make several
assumptions. First, the laser waves are considered as pure plane
waves. The atomic motion is then quantized in the direction
parallel to the laser beams. Second, we restrict our calculation
to the case of a constant Rabi frequency (square pulses). Third,
we assume the resonance condition is fulfilled. The Raman
interaction then couples the two states
$|a\rangle=|g_1,\overrightarrow{p}\rangle$ and
$|b\rangle=|g_2,\overrightarrow{p}+\hbar
\overrightarrow{k}_{eff}\rangle$ where $|g_1\rangle$ and
$|g_2\rangle$ are the two hyperfine levels of the ground state,
$\overrightarrow{p}$ is the atomic momentum,
$\overrightarrow{k}_{eff}$ is the difference between the wave
vectors of the two lasers.

We develop the atomic wave function on the basis set
$\{|a\rangle,|b\rangle\}$ so that
$|\Psi(t)\rangle=C_a(t)|a\rangle+C_b(t)|b\rangle$, and choose the
initial state to be $|\Psi(t_i)\rangle=|\Psi_i\rangle=|a\rangle$.
At the output of the interferometer, the transition probability is
given by $P=|C_b(t_f)|^2$, where $t_f=t_i+2T+4\tau_R$. The
evolution of $C_a$ and $C_b$ from $t_i$ to $t_f$ is given by
\begin{equation}
\left(
\begin{array}{c}
C_a(t_f) \\
C_b(t_f)
\end{array}
\right) = M \left(
\begin{array}{c}
C_a(t_i) \\
C_b(t_i)
\end{array}\right)
 \label{eq:Cevol}
\end{equation}
where $M$ is the evolution matrix through the whole
interferometer. Solving the Schr\"odinger equation gives the
evolution matrix during a Raman pulse\,\cite{Moler92}, from time
$t_0$ to time $t$:

\begin{equation}
M_p(t_0,t,\Omega_R,\phi)=\left(
\begin{array}{cc}
e^{-i\omega_a(t-t_0)}\text{cos}(\frac{\Omega_R}{2}(t-t_0)) & -ie^{-i\omega_a(t-t_0)}e^{i(\omega_Lt_0+\phi)}\text{sin}(\frac{\Omega_R}{2}(t-t_0))\\
-ie^{-i\omega_b(t-t_0)}e^{-i(\omega_Lt_0+\phi)}\text{sin}(\frac{\Omega_R}{2}(t-t_0))
& e^{-i\omega_b(t-t_0)}\text{cos}(\frac{\Omega_R}{2}(t-t_0))
\end{array}
\right) \label{eq:Mevol}
\end{equation}
where $\Omega_R/2\pi$ is the Rabi frequency and $\omega_L$, the effective frequency, is the
frequency difference between the two lasers, $\omega_L=\omega_2-\omega_1$. Setting $\Omega_R=0$ in
$M_p(t_0,t,\Omega_R,\phi)$ gives the free evolution matrix, which determines the evolution between
the pulses. The evolution matrix for the full evolution is obtained by taking the product of
several matrices. When $t$ occurs during the $i-th$ laser pulse, we split the evolution matrix of
this pulse at time $t$ into two successive matrices, the first one with $\phi_i$, and the second
one with $\phi =\phi_i+\delta \phi$.

Finally, we choose the time origin at the middle of the second
Raman pulse. We thus have $t_i=-(T+2\tau_R)$ and $t_f=T+2\tau_R$.
We then calculate the change in the transition probability for a
infinitesimally small phase jump at any time t during the
interferometer, and deduce $g(t)$. It is an odd function, whose
expression is given here for $t>0$:

\begin{equation}
\label{biggeq} g(t)=\left\{\begin{array}{ll}
 \sin(\Omega_R t) & 0<t<\tau_R \\
 1 & \tau_R <t<T+\tau_R\\
 -\sin(\Omega_R (T-t)) & T+\tau_R<t<T+2\tau_R\\
\end{array}\right.
\end{equation}

When the phase jump occurs outside the interferometer, the change
in the transition probability is null, so that $g(t)=0$ for
$|t|>T+2\tau_R$.

In order to validate this calculation, we use the gyroscope
experiment to measure experimentally the sensitivity function.
About $10^8$ atoms from a background vapor are loaded in a 3D-MOT
within 125 ms, with 6 laser beams tuned to the red of the
$F=4\rightarrow F'=5$ transition at 852 nm. The atoms are then
launched upwards at $\sim 2.4$ m/s within 1 ms, and cooled down to
an effective temperature of $\sim 2.4\mu$K. After launch, the
atoms are prepared into the $\left|F=3,m_F=0\right\rangle$ state
using a combination of microwave and laser pulses : they first
enter a selection cavity tuned to the
$\left|F=4,m_F=0\right\rangle\rightarrow\left|F=3,m_F=0\right\rangle$
transition. The atoms left in the $F=4$ state are pushed away by a
laser beam tuned to the $F=4\rightarrow F'=5$ transition, 11 cm
above the selection cavity. The selected atoms then reach the
apogee 245 ms after the launch, where they experience three
interferometer pulses of duration $\tau_R-2\tau_R-\tau_R$ with
$\tau_R=20~\mu$s separated in time by $T=4.97$ ms. The number of
atoms $N_{F=3}$ and $N_{F=4}$ are finally measured by detecting
the fluorescence induced by a pair of laser beams located 7 cm
below the apogee. From these measurements, we deduce the
transition probability $N_{F=4}/(N_{F=3}+N_{F=4})$. The total
number of detected atoms is about $10^5$. The repetition rate of
the experiment is 2 Hz.

\begin{figure}[h]
\centering
\includegraphics[width=5in]{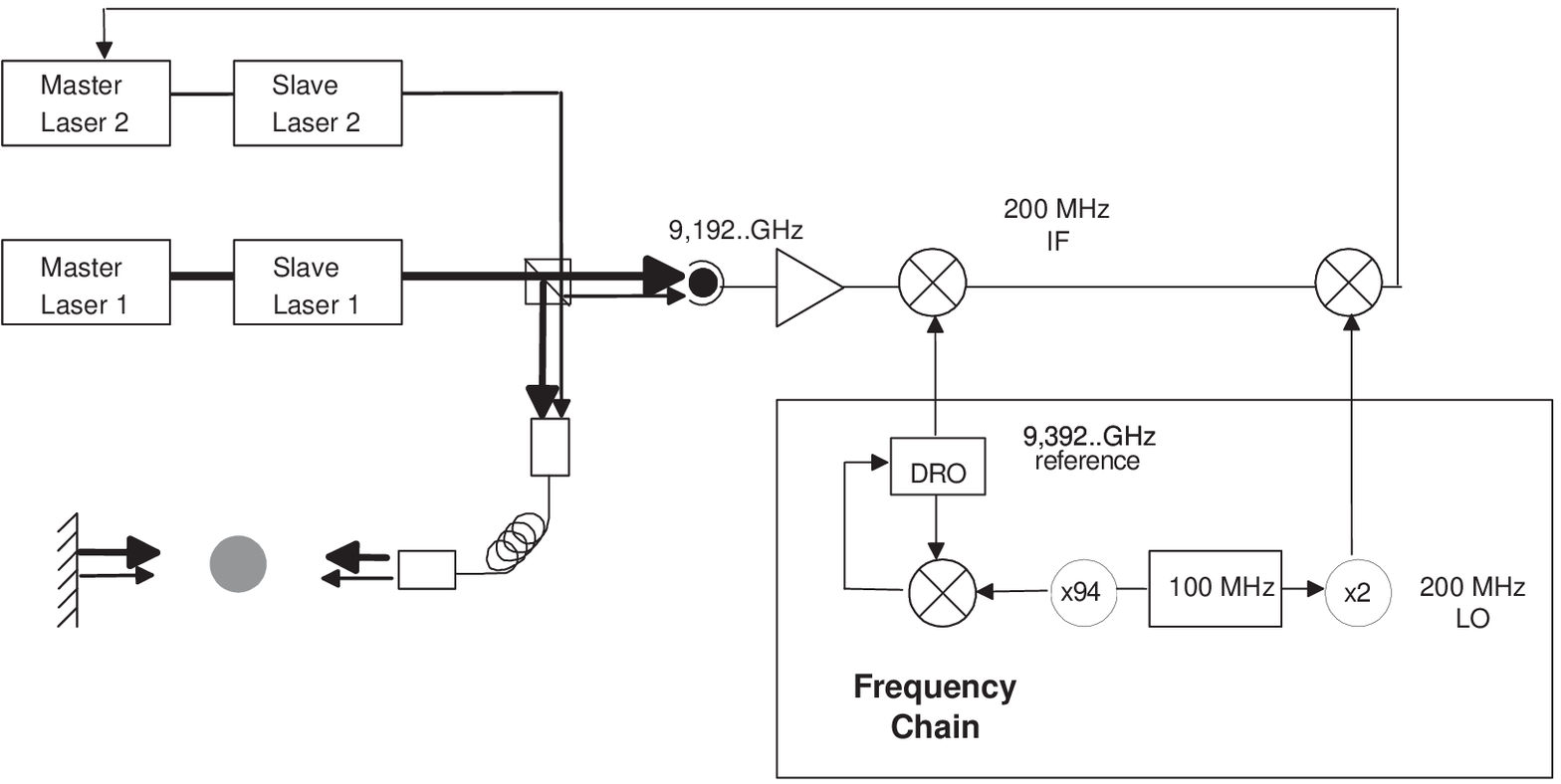}
\caption{Principle of the laser phase-lock: the beatnote at $9.192$ GHz between the two Raman
lasers is observed on a fast response photodetector. After amplification, this beatnote is mixed
with the reference frequency at $9.392$ GHz from the frequency chain, to obtain a signal at $200$
MHz. This signal is compared with the reference frequency at $200$ MHz from the same frequency
chain to get an error signal. This error signal is then processed and sent to the current of the
laser and to the PZT that controls the laser cavity length.} \label{raman}
\end{figure}

The set-up for the generation of the two Raman laser beams is
displayed in figure \ref{raman}. Two slave diode lasers of 150 mW
output power are injected with extended cavity diode lasers. The
polarizations of the slave diodes output beams are made orthogonal
so that the two beams can be combined onto a polarization beam
splitter cube. The light at this cube is then split in two
distinct unbalanced paths.

On the first path, most of the power of each beam is sent through
an optical fiber to the vacuum chamber. The two beams are then
collimated with an objective attached onto the chamber (waist
$w_0=15$ mm). They enter together through a viewport, cross the
atomic cloud, and are finally retroreflected by a mirror fixed
outside the vacuum chamber. In this geometry, four laser beams are
actually sent onto the atoms, which interact with only two of
them, because of selection rules and resonance conditions. The
interferometer can also be operated with co-propagating Raman
laser beams by simply blocking the light in front of the
retroreflecting mirror. A remarkable feature of this experiment is
that the three interferometer pulses are realized by this single
pair of Raman lasers that is turned on and off three times, the
middle pulse being at the top of the atoms' trajectory. For all
the measurements described in this article, the Raman lasers are
used in the $co-propagating$ configuration. The interferometer is
then no longer sensitive to inertial forces, but remains sensitive
to the relative phase of the Raman lasers. Moreover, as such Raman
transitions are not velocity selective, more atoms contribute to
the signal. All this allows us to reach a good signal to noise
ratio of 150 per shot.

The second path is used to control the Raman lasers phase
difference, which needs to be locked \cite{Santarelli94} onto the
phase of a very stable microwave oscillator. The phase lock loop
scheme is also displayed in figure\,\ref{raman}. The frequency
difference is measured by a fast photodetector, which detects a
beatnote at 9.192 GHz. This signal is then mixed with the signal
of a Dielectric Resonator Oscillator (DRO) tuned at 9.392 GHz. The
DRO itself is phase locked onto the 94th harmonics of a very
stable 100 MHz quartz. The output of the mixer (IF) is 200 MHz. A
local oscillator (LO) at 200 MHz is generated by doubling the same
100 MHz quartz. IF and LO are compared using a digital phase and
frequency detector, whose output is used as the error signal of
the phase-locked loop. The relative phase of the lasers is
stabilized by reacting on the current of one of the two diode
lasers, as well as on the voltage applied to the PZT that controls
the length of the extended cavity diode laser \cite{Santarelli94}.

\begin{figure}[h]
\centering
\includegraphics[width=5in]{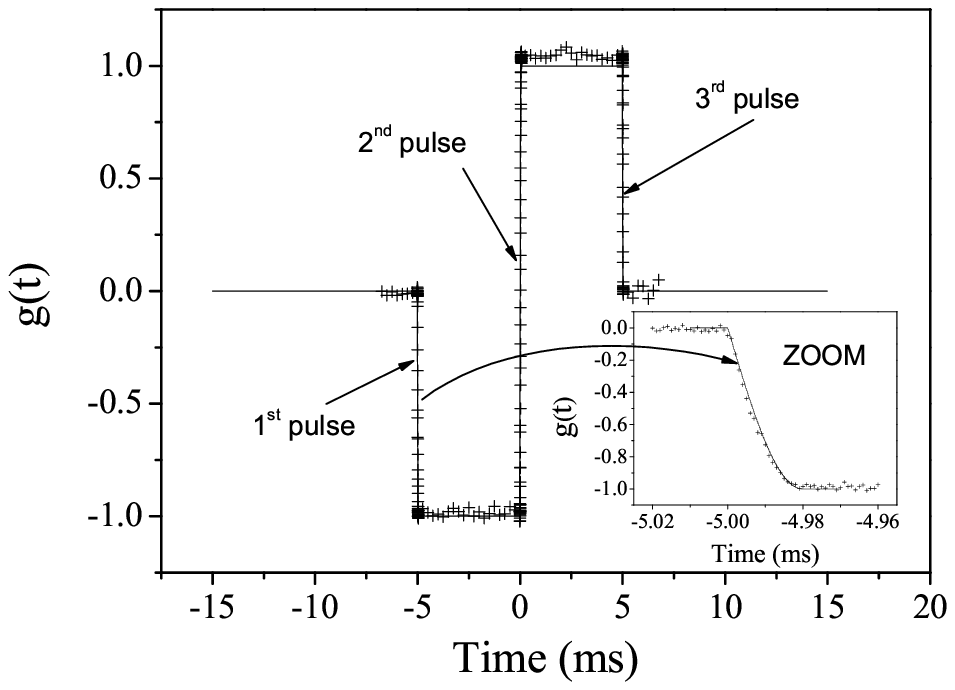}
\caption{The atomic sensitivity function $g(t)$ as a function of time, for a three pulses
interferometer with a Rabi frequency $\Omega_R=\frac{\pi}{2\tau_R}$. The theoretical calculation is
displayed in solid line and the experimental measurement with crosses. A zoom is made on the first
pulse.} \label{sensitivityplottotal}
\end{figure}

To measure $g(t)$, a small phase step of $\delta \phi=0.107$ rad
is applied at time $t$ on the local oscillator. The phase lock
loop copies this phase step onto the Raman phase within a fraction
of $\mu s$, which is much shorter than the Raman pulse duration of
$\tau_R=20~\mu s$. Finally we measured the transition probability
as a function of $t$ and deduced the sensitivity function. We
display in figure \ref{sensitivityplottotal} the measurement of
the sensitivity function compared with the theoretical
calculation. We also realized a precise measurement during each
pulse and clearly obtained the predicted sinusoidal rise of the
sensitivity function.

For a better agreement of the experimental data with the
theoretical calculation, the data are normalized to take into
account the interferometer's contrast, which was measured to be
$78\%$. This reduction in the contrast with respect to $100\%$ is
due to the combined effect of inhomogeneous Rabi frequencies
between the atoms, and unbalanced Rabi frequencies between the
pulses. Indeed, the atomic cloud size of $8$ mm is not negligible
with respect to the size of the single pair of Raman gaussian
beams, $w_0=15$\,mm. Atoms at both sides of the atomic cloud will
not see the same intensity, inducing variable transfer efficiency
of the Raman transitions. Moreover, the cloud moves by about $3$
mm between the first and the last pulse. In order for the cloud to
explore only the central part of the gaussian beams, we choose a
rather small interaction time of $T=4.97$ ms with respect to the
maximum interaction time possible of $T=40$ ms. Still, the
quantitative agreement is not perfect. One especially observes a
significant asymmetry of the sensitivity function, which remains
to be explained. A full numerical simulation could help in
understanding the effect of the experimental imperfections.

\section{Transfer Function of the interferometer\label{sec:transfer}}

From the sensitivity function, we can now evaluate the
fluctuations of the interferometric phase $\Phi$ for an arbitrary
Raman phase noise $\phi (t)$ on the lasers
\begin{equation}
\delta\Phi=\int_{-\infty}^{+\infty}g(t)d\phi(t)=\int_{-\infty}^{+\infty}g(t)\frac{d\phi(t)}{dt}dt.
\end{equation}
The transfer function of the interferometer can be obtained by
calculating the response of the interferometer phase $\Phi$ to a
sinusoidal modulation of the Raman phase, given by $\phi(t)=A_0
\rm{cos}(\omega_0 t+\psi)$. We find $\delta\Phi=A_0\omega_0
Im(G(\omega_0))\rm{cos}(\psi)$, where $G$ is the Fourier transform
of the sensitivity function.
\begin{equation}
G(\omega)= \int_{-\infty}^{+\infty}e^{-i\omega t}g(t)dt
\label{eq:G1}
\end{equation}

When averaging over a random distribution of the modulation phase
$\psi$, the rms value of the interferometer phase is $\delta
\Phi^{rms}=|A_0\omega_0 G(\omega_0)|$. The transfer function is
thus given by $H(\omega)=\omega G(\omega)$. If we now assume
uncorrelated Raman phase noise between successive measurements,
the rms standard deviation of the interferometric phase noise
$\sigma^{rms}_{\Phi}$ is given by:
\begin{equation}
\label{eq:fourier}
(\sigma^{rms}_{\Phi})^2=\int_{0}^{+\infty}|H(\omega)|^2
S_{\phi}(\omega)d\omega
\end{equation}
where $S_{\phi}(\omega)$ is the power spectral density of the
Raman phase.

We calculate the Fourier transform of the sensitivity function and
find:
\begin{equation}
G(\omega)=\frac{4i\Omega_R}{\omega^2-\Omega_R^2}\sin(\frac{\omega(T+2\tau_R)}{2})
(\cos(\frac{\omega(T+2\tau_R)}{2})+\frac{\Omega_R}{\omega}\sin(\frac{\omega
T}{2})) \label{eq:G3}
\end{equation}

At low frequency, where $\omega<<\Omega_R$, the sensitivity
function can be approximated by
\begin{equation}
G(\omega)=-\frac{4i}{\omega}\sin^2(\omega  T/2) \label{eq:G4}
\end{equation}

\begin{figure}[h]
\centering
\includegraphics[width=5in]{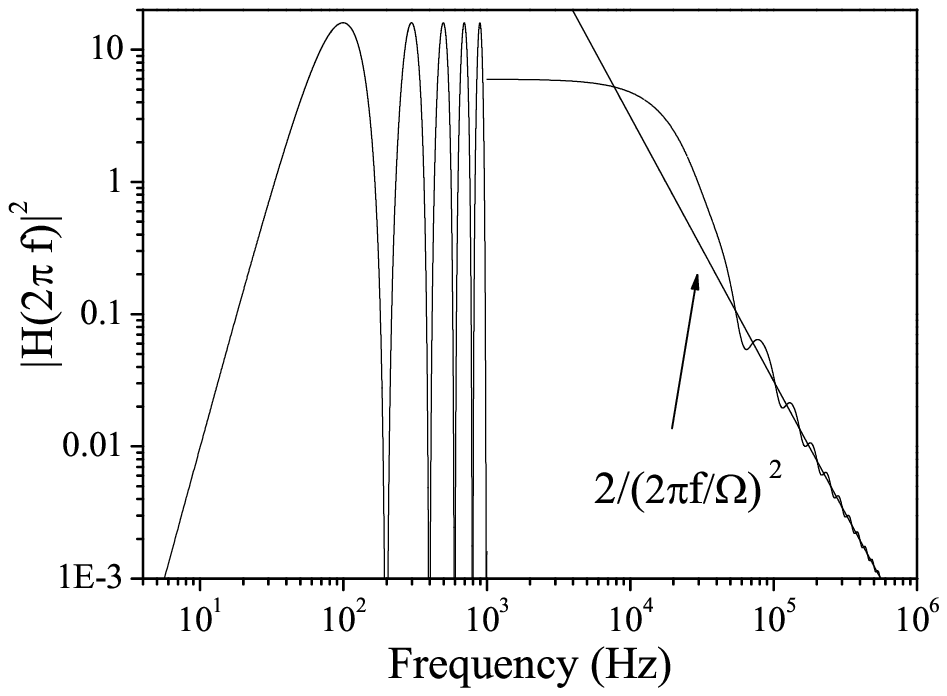}
\caption{Calculated weighting function for the Raman phase noise as a function of frequency. Below
1 kHz, the exact weighting function is displayed. It shows an oscillation with a period frequency
of $\delta f=\frac{1}{T+2\tau}$. Above 1 kHz only the mean value of the weighting function over
$\delta f$ is displayed. The weighting function acts as a first order low pass filter, with an
effective cutoff frequency of $f_0=\frac{\sqrt{3}}{3}\frac{\Omega_R}{2\pi}$} \label{ponderation}
\end{figure}

The weighting function $|H(2\pi f)|^{2}$ versus the frequency $f$
is displayed in figure \ref{ponderation}. It has two important
features: the first one is an oscillating behavior at a frequency
given by $1/(T+2\tau_R)$, leading to zeros at frequencies given by
$f_k=\frac{k}{T+2\tau_R}$. The second is a low pass first order
filtering due to the finite duration of the Raman pulses, with an
effective cutoff frequency $f_0$, given by
$f_0=\frac{\sqrt{3}}{3}\frac{\Omega_R}{2\pi}$. Above 1 kHz only
the mean value over one oscillation is displayed on the figure.

\begin{figure}[h]
\centering
\includegraphics[width=5in]{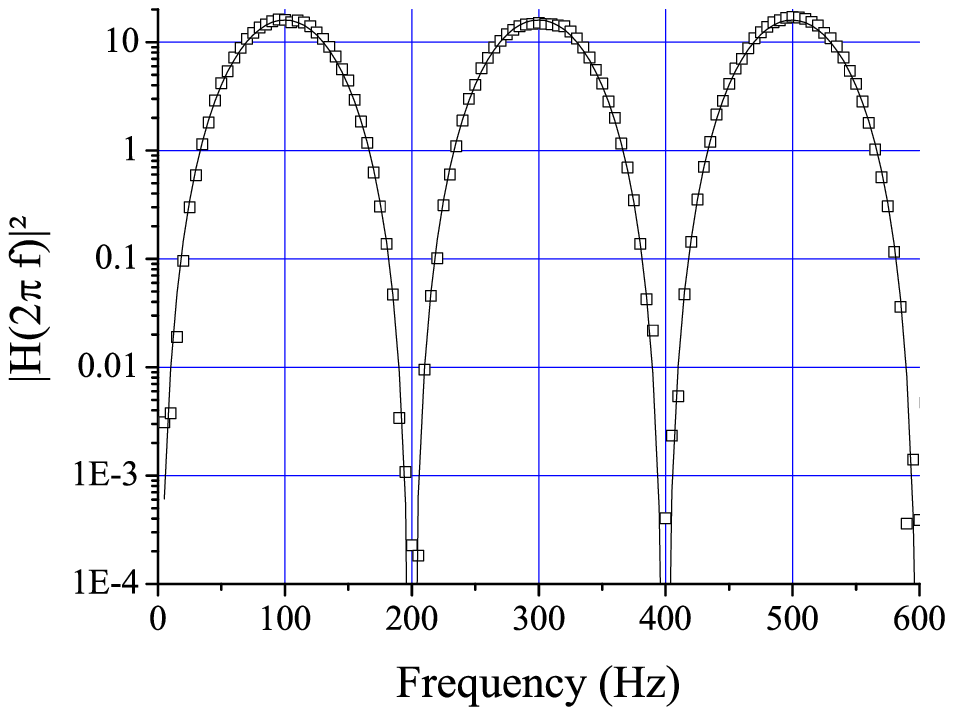}
\caption{The phase noise weighting function
  $|H(2\pi f)^2|$ for $T=4.97$\,ms and $\tau_R=20\,\rm{\mu s}$, at low frequency. The
  theoretical calculation is displayed in solid line and the
  experimental results in squares. We clearly see the oscillating
  behavior of the weighting function and the experimental
  measurement are in good agreement with the theoretical
  calculation.} \label{weightfct2}
\end{figure}

\begin{figure}[h]
\centering
\includegraphics[width=5in]{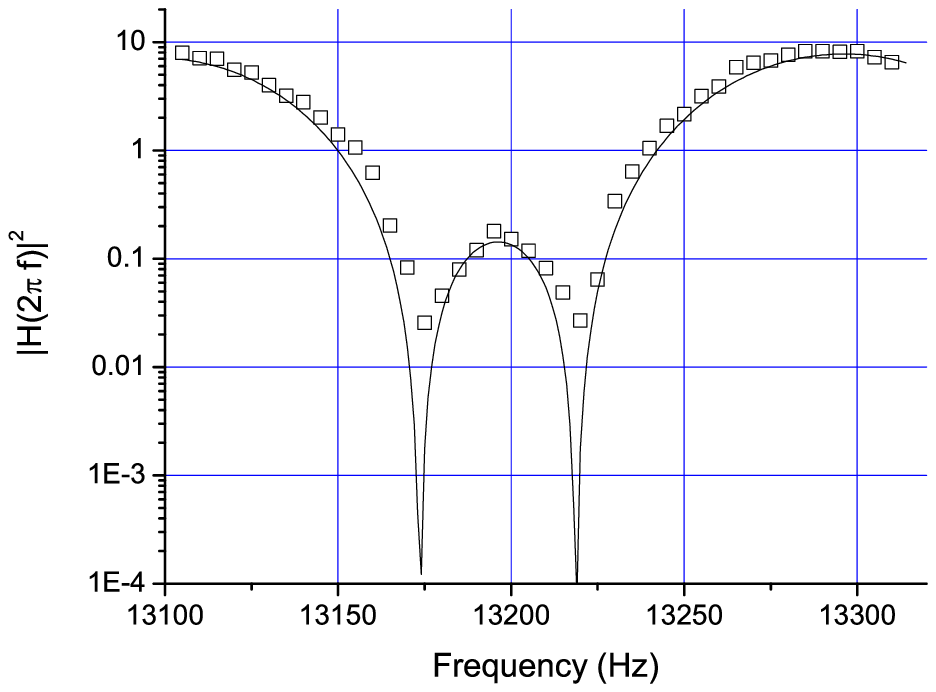}
\caption{The phase noise weighting function
  $|H(2\pi f)^2|$ for $T=4.97$\,ms and $\tau_R=20\,\rm{\mu s}$,displayed near the Rabi frequency. The
  theoretical calculation is displayed in solid line and the
  experimental results in squares. We identified the zero multiple
  of $\frac{1}{T+2\tau}$ and observed experimentally both zeros with a good agreement with theory.} \label{weightfct3}
\end{figure}

In order to measure the transfer function, a phase modulation $A_m
\rm{cos} (2\pi f_m t+\psi)$ is applied on the Raman phase,
triggered on the first Raman pulse. The interferometric phase
variation is then recorded as a function of $f_m$. We then repeat
the measurements for the phase modulation in quadrature $A_m
\rm{sin} (2\pi f_m t+\psi)$. From the quadratic sum of these
measurement, we extract $H(2\pi f_m)^2$. The weighting function
was first measured at low frequency. The results, displayed in
figure \ref{weightfct2} together with the theoretical value,
clearly demonstrate the oscillating behavior of the weighting
function. Figure \ref{weightfct3} displays the measurements
performed slightly above the cutoff frequency, and shows two
zeros. The first one corresponds to a frequency multiple of
$1/(T+2\tau)$. The second one is a zero of the last factor of
equation \ref{eq:G3}. Its position depends critically on the value
of the Rabi frequency.

When comparing the data with the calculation, the experimental
imperfections already mentioned have to be accounted for. An
effective Rabi frequency $\Omega_{eff}$ can be defined by the
relation $\Omega_{eff}\tau_0=\pi$, where $\tau_0$ is the duration
of the single pulse, performed at the center of the gaussian Raman
beams, that optimizes the transition probability. For homogeneous
Raman beams, this pulse would be a $\pi$ pulse. This effective
Rabi frequency is measured with an uncertainty of about 1 \%. It
had to be corrected by only 1.5 \% in order for the theoretical
and experimental positions of the second zero to match. The
excellent agreement between the theoretical and experimental
curves validate our model.

\section{Link between the sensitivity function and the sensitivity of the interferometer}

The sensitivity of the interferometer is characterized by the
Allan variance of the interferometric phase fluctuations,
$\sigma^{2}_{\Phi}(\tau)$, defined as

\begin{eqnarray}
 \sigma_{\Phi}^{2}(\tau)&=&\frac{1}{2}\langle(\bar{\delta \Phi}_{k+1}-\bar{\delta \Phi}_{k})^{2}\rangle \\
&=&\frac{1}{2}\lim_{n\rightarrow \infty}\left\{
 \frac{1}{n}\sum_{k=1}^{n}(\bar{\delta \Phi}_{k+1}-\bar{\delta \Phi}_{k})^{2}\right\}.\label{eq:variance_allan}
 \end{eqnarray}

where $\bar{\delta \Phi}_{k}$ is the average value of $\delta
\Phi$ over the interval $[t_{k},t_{k+1}]$ of duration $\tau$. The
Allan variance is equal, within a factor of two, to the variance
of the differences in the successive average values $\bar{\delta
\Phi}_{k}$ of the interferometric phase. Our interferometer being
operated sequentially at a rate $f_c=1/T_{\rm{c}}$, $\tau$ is a
multiple of $T_c$ : $\tau=m T_c$. Without loosing generality, we
can choose $t_{k}=-T_c/2+k m T_c$. The average value $\bar{\delta
\Phi}_{k}$ can now be expressed as

\begin{eqnarray}
 \bar{\delta \Phi}_{k}&=&\frac{1}{m}\sum_{i=1}^{m}\delta \Phi_{i}=\frac{1}{m}\sum_{i=1}^{m}\int_{t_{k}+(i-1)T_{\rm{c}}}^{t_{k}+iT_{\rm{c}}}g(t-t_{k}-(i-1)T_{\rm{c}}-T_{\rm{c}}/2)\frac{d\phi}{dt}~dt\\
 &=&\frac{1}{m}\int_{t_{k}}^{t_{k+1}}g_k(t)\frac{d\phi}{dt}~dt
\end{eqnarray}
where $g_k(t)=\sum_{i=1}^{m}g(t-k m T_{\rm{c}}-(i-1)T_{\rm{c}})$.
The difference between successive average values is then given by
\begin{equation}
 \bar{\delta \Phi}_{k+1}-\bar{\delta \Phi}_{k}=\frac{1}{m}\int_{-\infty}^{+\infty}(g_{k+1}(t)-g_{k}(t))\frac{d\phi}{dt}~dt
\end{equation}

For long enough averaging times, the fluctuations of the
successive averages are not correlated and the Allan variance is
given by
\begin{equation}
 \sigma_{\Phi}^{2}(\tau)=\frac{1}{2}\frac{1}{m^2}\int_{0}^{+\infty}
|G_m(\omega)|^2 \omega^2 S_{\phi}(\omega)d\omega
\end{equation}
where $G_m$ is the Fourier transform of the function
$g_{k+1}(t)-g_{k}(t)$. After a few algebra, we find for the
squared modulus of $G_m$ the following expression
\begin{equation}
|G_m(\omega)|^2=4\frac{\rm{sin}^4(\omega m
T_c/2)}{\rm{sin}^2(\omega T_c/2)}|G(\omega)|^2
\end{equation}
When $\tau \rightarrow \infty$, $|G_m(\omega)|^2 \sim
\frac{2m}{T_c} \sum_{j=-\infty}^{\infty}\delta(\omega-j2\pi
f_{\rm{c}})|G(\omega)|^2$. Thus for large averaging times $\tau$,
the Allan variance of the interferometric phase is given by
\begin{equation}
\label{Dick} \sigma^{2}_{\Phi}(\tau)={1\over
\tau}\sum_{n=1}^{\infty}|H(2\pi n f_{\rm{c}})|^2
        S_{\phi}({2\pi n f_{\rm{c}}})
\end{equation}
Equation \ref{Dick} shows that the sensitivity of the
interferometer is limited by an aliasing phenomenon similar to the
Dick effect in atomic clocks\,\cite{Dick87} : only the phase noise
at multiple of the cycling frequency appear in the Allan variance,
weighted by the Fourier components of the transfer function.

Let's examine now the case of white Raman phase noise :
$S_{\phi}(\omega)=S_{\phi}^{\rm{0}}$. The interferometer
sensitivity is given by:
\begin{equation}
\label{whiteeq1} \sigma^{2}_{\Phi}(\tau)=(\frac{\pi}{2})^2
\frac{S_{\phi}^{\rm{0}}}{\tau}\frac{T_{\rm{c}}}{\tau_R}
\end{equation}
In that case, the sensitivity of the interferometer depend not
only on the Raman phase noise spectral density but also on the
pulse duration $\tau_R$. For a better sensitivity, one should use
the largest pulse duration as possible. But, as the Raman
transitions are velocity selective, a very long pulse will reduce
the number of useful atoms. This increases the detection noise
contribution, so that there is an optimum value of $\tau_R$ that
depends on the experimental parameters. In the case of the
gyroscope, the optimum was found to be $\tau_R=20\,\mu$s.

To reach a good sensitivity, the Raman phase needs to be locked to
the phase of a very stable microwave oscillator (whose frequency
is 6.834 GHz for $^{87}$Rb and 9.192 GHz for $^{133}$Cs). This
oscillator can be generated by a frequency chain, where low phase
noise quartz performances are transposed in the microwave domain.
At low frequencies ($f<10-100$ Hz), the phase noise spectral
density of such an oscillator is usually well approximated by a
$1/f^3$ power law (flicker noise), whereas at high frequency
($f>1$ kHz), it is independent of the frequency (white noise).
Using equation \ref{Dick} and the typical parameters of our
experiments ($\tau_R=20\,\rm{\mu s}$ and $T=50$ ms), we can
calculate the phase noise spectral density required to achieve an
interferometric phase fluctuation of 1 mrad per shot. This is
equivalent to the quantum projection noise limit for $10^6$
detected atoms. The flicker noise of the microwave oscillator
should be lower than $-53\,\rm{dB.rad}^2.\rm{Hz}^{-1}$ at 1 Hz
from the carrier frequency, and its white noise below
$-111\,\rm{dB.rad}^2.\rm{Hz}^{-1}$. Unfortunately, there exists no
quartz oscillator combining these two levels of performance. Thus,
we plan to lock a SC Premium 100 MHz oscillator (from Wenzel
Company) onto a low flicker noise 5 MHz Blue Top oscillator
(Wenzel). From the specifications of these quartz, we calculate a
contribution of 1.2 mrad to the interferometric phase noise.

\begin{figure}[h]
\centering
\includegraphics[width=5in]{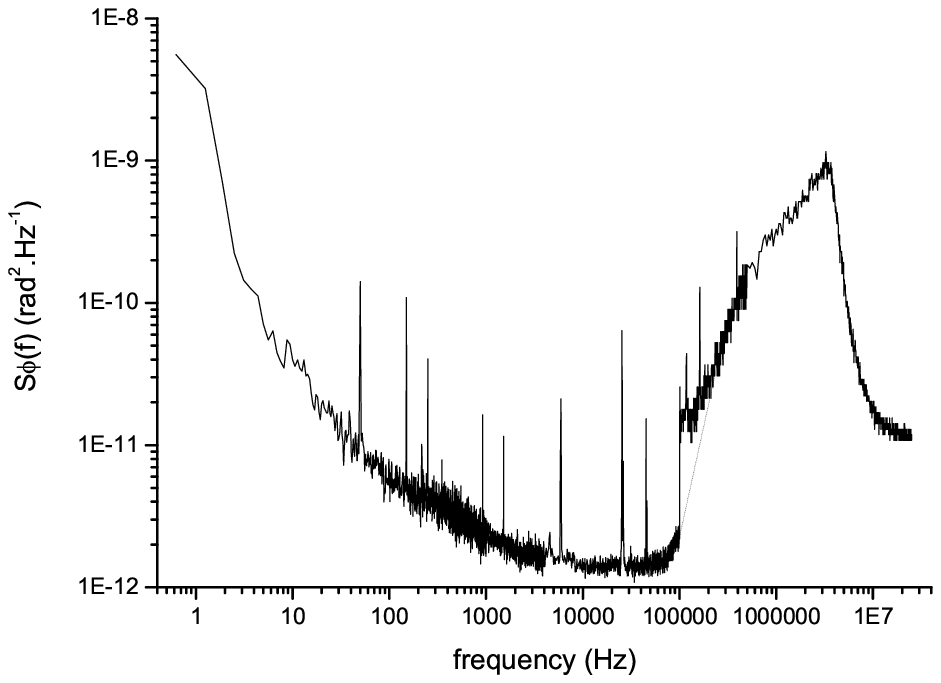}
\caption{Phase noise power spectral density between the two phase locked diode
  lasers. Up to 100 kHz, we display the residual noise of the phaselock
  loop, obtained by measuring the phase noise of the
  demodulated beatnote on a Fast Fourier Transform analyzer. There, the phase noise of the reference
  oscillator is rejected. Above 100 kHz, we display the phase noise measured directly on the beatnote
  observed onto a spectrum analyzer. In this case, the reference
  oscillator phase noise limits the Raman phase noise to $1.5\times 10^{-11}
  \rm{rad}^2.\rm{Hz}^{-1}$. In doted line is displayed an extrapolation of
  the phase noise due to the phase-lock loop alone between 100 kHz
  and 300 kHz.} \label{phaselock}
\end{figure}

Phase fluctuations also arise from residual noise in the
servo-lock loop. We have measured experimentally the residual
phase noise power spectral density of a phase lock system
analogous to the one described in figure \ref{raman}. This system
has been developed for phase locking the Raman lasers of the
gravimeter experiment. The measurement was performed by mixing IF
and LO onto an independent RF mixer, whose output phase
fluctuations was analyzed onto a Fast Fourier Transform analyzer.
The result of the measurement is displayed on figure
\ref{phaselock}. At low frequencies, below 100 Hz, the phase noise
of our phaselock system lies well below the required flicker
noise. After a few kHz, it reaches a plateau of
$-119\,\rm{dB.rad}^2.\rm{Hz}^{-1}$. The amplitude of this residual
noise is not limited by the gain of the servo loop. Above 60 kHz,
it increases up to $-90\,\rm{dB.rad}^2.\rm{Hz}^{-1}$ at $3.5$ MHz,
which is the bandwidth of our servo lock loop. Using equation
\ref{Dick}, we evaluated to $0.72$ mrad its contribution to the
interferometer's phase noise.

Other sources of noise are expected to contribute, which haven't
been investigated here : noise of the fast photodetector, phase
noise due to the propagation of the Raman beams in free space and
in optical fibers \cite{Yver03}.

\section{The case of parasitic vibrations\label{sec:vibrations}}

The same formalism can be used to evaluate the degradation of the
sensitivity caused by parasitic vibrations in the usual case of
counter-propagating Raman beams. As the two laser beams are first
overlapped before being sent onto the atoms, their phase
difference is mostly affected by the movements of a single optical
element, the mirror that finally retro-reflects them.

A displacement of this mirror by $\delta z$ induces a Raman phase
shift of $k_{eff}\delta z$. The sensitivity of the interferometer
is then given by
\begin{equation}
\label{Vibeq} \sigma^{2}_{\Phi}(\tau)={k_{eff}^2\over
\tau}\sum_{n=1}^{\infty}|H(2\pi n f_{\rm{c}})|^2
        S_{z}(2\pi n f_{\rm{c}})
\end{equation}
where $S_{z}(\omega)$ is the power spectral density of position
noise. Introducing the power spectral density of acceleration
noise $S_{z}(\omega)$, the previous equation can be written

\begin{equation}
\label{Vibeq2} \sigma^{2}_{\Phi}(\tau)={k_{eff}^2\over
\tau}\sum_{n=1}^{\infty}
        \frac{|H(2\pi n f_{\rm{c}})|^2}{(2\pi n f_{\rm{c}})^4}
        S_{a}(2\pi n f_{\rm{c}})
\end{equation}
It is important to note here that the acceleration noise is
severely filtered by the transfer function for acceleration which
decreases as $1/f^4$.

In the case of white acceleration noise $S_{a}$, and to first
order in $\tau_R/T$, the limit on the sensitivity of the
interferometer is given by :

\begin{equation}
\label{whiteeq} \sigma^{2}_{\Phi}(\tau)=\frac{k_{eff}^2
T^4}{2}\left(\frac{2T_{\rm{c}}}{3T}-1\right)\frac{S_{a}}{\tau}\
\end{equation}

To put this into numbers, we now calculate the requirements on the
acceleration noise of the retroreflecting mirror in order to reach
a sensitivity of 1 mrad per shot. For the typical parameters of
our gravimeter, the amplitude noise should lie below
$10^{-8}\,\rm{m.s^{-2}.Hz^{-1/2}}$. The typical amplitude of the
vibration noise measured on the lab floor is
$2\times10^{-7}\,\rm{m.s^{-2}.Hz^{-1/2}}$ at 1 Hz and rises up to
about $5\times10^{-5}\,\rm{m.s^{-2}.Hz^{-1/2}}$ at 10 Hz. This
vibration noise can be lowered to a few
$10^{-7}\,\rm{m.s^{-2}.Hz^{-1/2}}$ in the 1 to 100 Hz frequency
band with a passive isolation platform. To fill the gap and cancel
the effect of vibrations, one could use the method proposed in
\cite{Yver03}, which consists in measuring the vibrations of the
mirror with a very low noise seismometer and compensate the
fluctuations of the position of the mirror by reacting on the
Raman lasers phase difference.

\section{Conclusion\label{sec:conclusion}}

We have here calculated and experimentally measured the
sensitivity function of a three pulses atomic interferometer. This
enables us to determine the influence of the Raman phase noise, as
well as of parasitic vibrations, on the noise on the
interferometer phase. Reaching a 1 mrad shot to shot fluctuation
requires a very low phase noise frequency reference, an optimized
phase lock loop of the Raman lasers, together with a very low
level of parasitic vibrations. With our typical experimental
parameters, this would result in a sensitivity of
$4\times10^{-8}\,\rm{rad.s^{-1}.Hz^{-1/2}}$ for the gyroscope and
of $1.5\times10^{-8}\,\rm{m.s^{-2}.Hz^{-1/2}}$ for the gravimeter.

Improvements are still possible. The frequency reference could be
obtained from an ultra stable microwave oscillator, such as a
cryogenic sapphire oscillator \cite{Mann01}, whose phase noise
lies well below the best quartz available. Besides, the
requirements on the phase noise would be easier to achieve using
atoms with a lower hyperfine transition frequency, such as Na or
K. Trapping a very large initial number of atoms in the 3D-MOT
would enable a very drastic velocity selection. The duration of
the Raman pulses could then be significantly increased, which
makes the interferometer less sensitive to high frequency Raman
phase noise. The manipulation of the atoms can also be implemented
using Bragg pulses \cite{Rasel95,Giltner95}. The difference in the
frequencies of the two beams being much smaller, the requirements
on the relative phase stability is easy to achieve. In that case,
a different detection method needs to be implemented as atoms in
both exit ports of the interferometer are in the same internal
state. Using ultracold atoms with subrecoil temperature, atomic
wavepackets at the two exit ports can be spatially separated,
which allows for a simple detection based on absorption imaging.
Such an interferometer would benefit from the long interaction
times available in space to reach a very high sensitivity.

We also want to emphasize that the sensitivity function can also
be used to calculate the phase shifts arising from all possible
systematic effects, such as the light shifts, the magnetic field
gradients and the cold atom collisions.

\section*{Acknowledgment}

The authors would like to thank Andr{\'e} Clairon for fruitful
discussions and careful reading of the manuscript. This work was
supported in part by BNM, CNRS, DGA and CNES. BNM-SYRTE is Unit\'e
Associ\'ee au CNRS, UMR 8630.

\end{document}